\documentclass[useAMS,usenatbib,usegraphicx]{mn2e}
\usepackage {times}

\title[FPs of E+As and UV-excess early-type galaxies]
{The Fundamental Planes of E+A galaxies and GALEX UV-excess
early-type galaxies: Revealing their intimate connection}
\author[Y. Choi, T. Goto and S.-J. Yoon]{Yumi Choi$^{1}$,
Tomotsugu Goto$^{2,3}$
and Suk-Jin Yoon$^{1}$\thanks{E-mail: sjyoon@galaxy.yonsei.ac.kr} \\
$^{1}$Department of Astronomy and Centre for Space Astrophysics, Yonsei University, Seoul 120-749, Korea\\
$^{2}$Institute of Space and Astronautical Science, Japan Aerospace Exploration Agency,
      Sagamihara, Kanagawa 229-8510, Japan\\
$^{3}$Institute for Astronomy, University of Hawaii
2680 Woodlawn Drive, Honolulu, HI, 96822, USA\\}
\begin{document}

%\date{Accepted ? December ?. Received ? }

%\pagerange{\pageref{firstpage}--\pageref{lastpage}} \pubyear{2008}

\maketitle

\label{firstpage}

\begin{abstract}
Strong Balmer absorption lines and the lack of H$\alpha$ and
[OII] emission lines signify that E+A galaxies are post-starburst systems.
Recent studies suggest that E+As may undergo the transition from
the `blue cloud' to the `red sequence' and eventually migrate to red sequence
early-type galaxies. An observational validation of this scenario
is to identify the intervening galaxy population between E+As and the red-sequence.
Motivated by recent findings with Galaxy Evolution Explorer (GALEX) that an unexpectedly large fraction of
early-type galaxies exhibit UV-excess (i.e. blue UV -- optical colours)
as a sign of recent star formation (RSF), we investigate the possible
connection of the UV-excess galaxies to E+As. In particular,
we examine the Fundamental Plane (FP) scaling relations of the currently
largest sample of $\sim$1,000 E+As selected from the SDSS and $\sim$20,000
morphologically-selected SDSS early-type galaxies with GALEX UV data.
The FP parameters, combined with stellar population indicators,
reveal a certain group of UV-excess early-types that bridges between
E+As and quiescent red galaxies. The newly identified galaxies
are the post-starburst systems characterized by UV-excess but no
H$\alpha$ emission. This is essentially a conceptual generalisation
of ``E+A'', in that the Balmer absorption line in the ``E+A" definition
is replaced with UV -- optical colours that are far more sensitive to
RSF than the Balmer lines. We refer to these UV-excess galaxies as
``E+a" galaxies (named after ``E+A"), which stands for elliptical (``E'')
galaxies with a minority of A-type (``a'') young stars. The species
are either (1) galaxies that experienced starbursts weaker than those
observed in E+As (1 $\sim$ 10 per cent of E+As, ``mild E+As'') or
(2) the products of passively evolved E+As after quenching star formation
quite a while ago ($\sim$ 1 Gyr, ``old E+As''). We suggest that the latter
type of E+a galaxies (i.e. old ``E+As'') represents the most recent arrival to the red sequence
in the final phase of the ¡°E+A¡± to ¡°red early-type¡± transition.
\end{abstract}


\section{Introduction}
The study of galaxy evolution began to flourish after \citet{dressler83}
discovered galaxies with unusual spectra that showed characteristics of both
elliptical galaxies and A-type stars, now referred to as ``E+A'' galaxies.
E+As are considered as post-starburst systems because of strong Balmer
absorption lines, sign of recent starbursts within $\sim$ 1 Gyr, and a
lack of [OII] and H$\alpha$ emission lines indicating a suddenly quenched
star formation \citep[e.g.][]{dressler83,poggianti99,goto04}.

How star formation in E+As has started and quenched is a topic that
has been studied by many.
One plausible explanation is that E+As result from some physical mechanisms
in cluster environments, such as ram-pressure stripping
\citep[e.g.][]{spitzer51,gunn72,fujita04}, galaxy-cluster tidal interaction
\citep{bryd90}, or galaxy harassment \citep[e.g.][]{moore96}.
However, the fact that a significant fraction of E+As are found in field
environments implies that cluster-related physical mechanisms are not enough to
explain the origin of all E+As \citep[e.g.][]{zabludoff96,blake04,goto05}.
Another possible explanation for the E+A phenomena is starburst galaxies with very thick
optical depth caused by dust obscuration \citep{couch87,smail99,poggianti00},
but radio and IR observations have detected little star formation obscured
by dust in E+As \citep{galaz00,miller01,goto04}.
Recently, \citet{goto05} showed that E+As have closer companions
than other galaxies do, providing evidence that a dynamical
merger\,/\,interaction could be the physical origin of field E+As.
This scenario is also supported by the dynamically disturbed morphologies
of E+As \citep{tran04,yang04,yamauchi05,liu07}. While there are some
plausible scenarios, the origin of E+As is still not well understood.

E+As are the key populations for understanding the galaxy evolution
because they are believed to be in a transition phase from a `blue cloud' to a `red sequence'.
Based on galaxy properties such as colour profiles and scaling relations,
\citet{yang08} suggest that E+As are the remnants of galaxy-galaxy
interactions/mergers and ultimately evolve into early-type galaxies.
The typical migration time from the gas-rich `blue cloud' to the gas-poor
`red sequence' is about 1.5 Gyr according to the UV properties of E+As
\citep{kaviraj07}.
UV light is useful for tracking E+A migration because it is very sensitive
to the young stellar population. The recent star formation (RSF) signature in UV
light continues about 1.5 Gyr \citep{yi05}. According to recent GALEX
observations, a significant fraction of early-type galaxies in the local Universe
have enhanced UV flux (i.e. blue UV -- optical colours) as a sign of young stellar population
\citep[e.g.][]{yoon04a,yi05,salim05,donas07,kaviraj07b}. The relatively weak RSF galaxies
might have arisen mainly through two channels: (a) the residual star formation in gas-poor
galaxies, or (b) the fading away of a young stellar population in galaxies that
underwent a violent starburst. UV-excess early-types
appear to be the most recent arrival at the red sequence in the final
phase of the blue cloud to red sequence transition, and may thus be linked to E+As.

Early-type galaxies can be described by three observational parameters:
their effective radius ($r_{e}$), effective mean
surface brightness ($\mu$) within $r_{e}$, and central velocity dispersion ($\sigma$).
These parameters are unified in a two-dimensional manifold, the Fundamental Plane (FP)
\citep{djor87,dressler87}. Even though the FP has intrinsic scatter \citep{jorgensen96},
the thin FP of early-type galaxies implies that there are well defined scaling
relations between their global parameters. The FP can be expressed such that
log $R_{e}$ = $\alpha$ log $\sigma$ + $\beta$ $\mu$ + $\gamma$,
where $R_{e}$ is the effective radius in kpc.
Supposed that early-type galaxies are virialized homologous systems with a
constant mass-to-light ratio, the virial theorem predicts $\alpha$ = 2.0
and $\beta$ = 0.4. However, the actual, measured coefficients are different from
the theoretical expectations (called the ``tilt'' of the FP).
The FP is a useful tool for exploring evolutionary connections among
various classes of galaxies \citep[e.g.][]{genzel01,tacconi02,lee07}.
The purpose of this paper is to investigate the evolutionary path from E+As to
quiescent galaxies via UV-excess galaxies by comparing their FPs.

The paper is organized as follows. A description of the sample used in this study
is given in \S\,2. We examine the FPs of our samples and
discuss the implication of our results in \S\,3. We will report an evolutionary
connection between E+As and UV-excess galaxies. Finally, \S\,4 gives the summary.

Except where stated otherwise, we adopt the $\Lambda$CDM cosmology with
$(h,\Omega_m,\Omega_\Lambda) = (0.7,0.3,0.7)$.
% End of introduction

\section[]{Sample}
\label{sec:sample}
\subsection{E+A galaxies}
This study uses the largest sample of 1,284 E+As selected from the SDSS DR6.
The selection criteria of the E+As are the same as \citet{goto07}, i.e.
H$\delta$ EW $>$ 4 \AA, [OII] EW $>$ --2.5 \AA, H$\alpha$ EW $>$ --3.0 \AA,
and a signal-to-noise ratio $>$ 10 pixel$^{-1}$ in the $r$-band.
Since velocity dispersions were measured for only 415 E+As in the SDSS
pipeline, we independently measure the velocity dispersions of 673,668
galaxies in the SDSS DR6.
Our method is similar to that utilized by the SDSS spectroscopic pipeline;
we first mask out possible emission line regions, and then fit the rest of
the spectra with a combination of eigen spectra with varying velocity dispersions.

\begin{figure}
\begin{center}
\includegraphics[width=9cm,height=9cm]{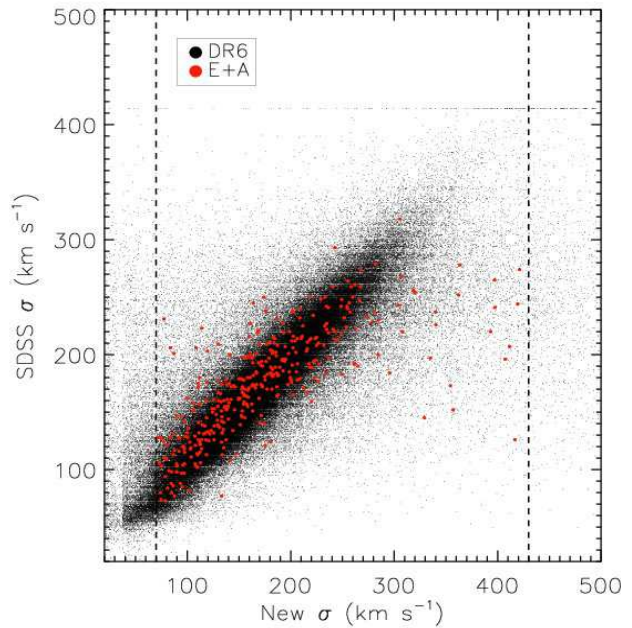}
\caption{SDSS pipeline velocity dispersions vs. our measurements for 673,668
 galaxies in the SDSS DR6 catalog. The two dashed vertical lines are the
 lower (70 km~s$^{-1}$) and the upper (430 km~s$^{-1}$) limits
 of the velocity dispersion in our sample.
 The instrumental dispersion of the SDSS spectrograph sets the
 lower limit (69 km~s$^{-1}$). The upper limit is given above which
 none of template spectra fits the observed ones for most of the galaxies.
 Red circles are the 377 E+As in common, satisfying 70 km~s$^{-1}$ $\le$ $\sigma$
 $\le$ 430 km~s$^{-1}$ in both measurements.}
\label{fig:vdisp}
\end{center}
\end{figure}

Figure~\ref{fig:vdisp} compares our velocity dispersion measurements with those
from the SDSS pipeline, showing reasonable one-to-one matches. Red circles are the
377 E+As in  common, satisfying 70 km~s$^{-1}$ $\le$ $\sigma$ $\le$ 430 km~s$^{-1}$ in both
measurements. We restrict our sample to galaxies with 70 km~s$^{-1}$ $\le$ $\sigma$
$\le$ 430 km~s$^{-1}$ because the instrumental dispersion of the SDSS spectrograph
is 69 km~s$^{-1}$ per pixel, and none of template spectra matches the observed ones
for most of the galaxies with $\sigma$ of 430 km~s$^{-1}$ or higher. Dashed vertical
lines mark the lower (70 km~s$^{-1}$) and upper (430 km~s$^{-1}$) limits on the
velocity dispersion of our sample. Among the 1,284 E+A candidates from the
SDSS DR6, we will use 1,021 E+As which satisfy the velocity dispersion criterion.
This is the largest and the most homogeneous E+A sample ever investigated.
%
% Velocity Dispersion Figure

\subsection{GALEX\,/\,SDSS early-type galaxies}
The GALEX\,/\,SDSS early-type galaxy sample is constructed as follows. First, we make use of
$\sim$ 180,000 morphologically-selected SDSS DR4plus early-type galaxies from
\citet{choi07}.
In this morphological classification method first presented by \citet{park05},
$u-r$ colour, radial gradient in $g-i$ colour, and concentration are the criteria for classification.
\citet{park05} estimated the completeness and reliability of this classification method
reaching about 91 per cent for SDSS galaxies brighter than $r$-band Petrosian mag, $r_{pet}$ $<$ 15.9
and about 88 per cent at $r_{pet}$ $<$ 17.5.
Then, we match the SDSS early-type galaxies with UV detections in the
2,241 GALEX fields from the GALEX DR4, imaged in the Medium Imaging Survey
(MIS), the Deep Imaging Survey (DIS), and Nearby Galaxy Survey (NGS) modes.
The cross-matching is performed with an angular matching radius of 3 arcsec.
For the multiple GALEX\,/\,SDSS matches within the given matching radius
($\sim$ 5 per cent of all matches), we select the UV source with the smallest angular
distance from the SDSS galaxy.
The final catalog consists of 23,539 early-type galaxies with near-UV (NUV)
information ($\sim$ 15 per cent of the entire early-type galaxy sample in Choi et al. 2007).

\begin{figure}
\begin{center}
\includegraphics[width=8.5cm,height=6.5cm]{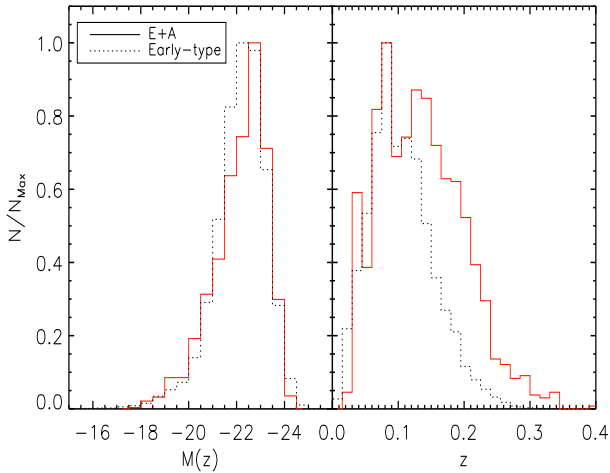}
\caption{Distributions of $z$-band absolute mag, $M_z$, and redshift, $z$, of the sample used in this study.
Solid lines are for E+A galaxies and dotted lines are for the early-type galaxies with GALEX UV information.}
\label{fig:zMzDist}
\end{center}
\end{figure}

To give basic information of our sample, Figure~\ref{fig:zMzDist} provides
the distributions of $z$-band absolute mag, $M_Z$, and redshift
of E+A and early-type galaxies. The early-type sample shows very similar $M_Z$ distribution to that of E+A,
suggesting their similar stellar mass ranges.
However, due to the GALEX UV detection limit, the smaller galaxies are selectively missing
in the higher redshift bins, resulting in early-types being confined to lower-redshift regime
compared to E+A. In order to avoid the selection bias, the comparative analysis between early-type
galaxies and E+As will be restricted to the early-type sample with $z$ $<$ 0.1. Also note that
our volume-limited early-type sample will have redshift ranging from 0.025 to 0.09.

%
% Distribution of z & M_z
%
% Hd vs. g-r / NUV-r Figure
\begin{figure}
\begin{center}
\includegraphics[width=9cm,height=12.5cm]{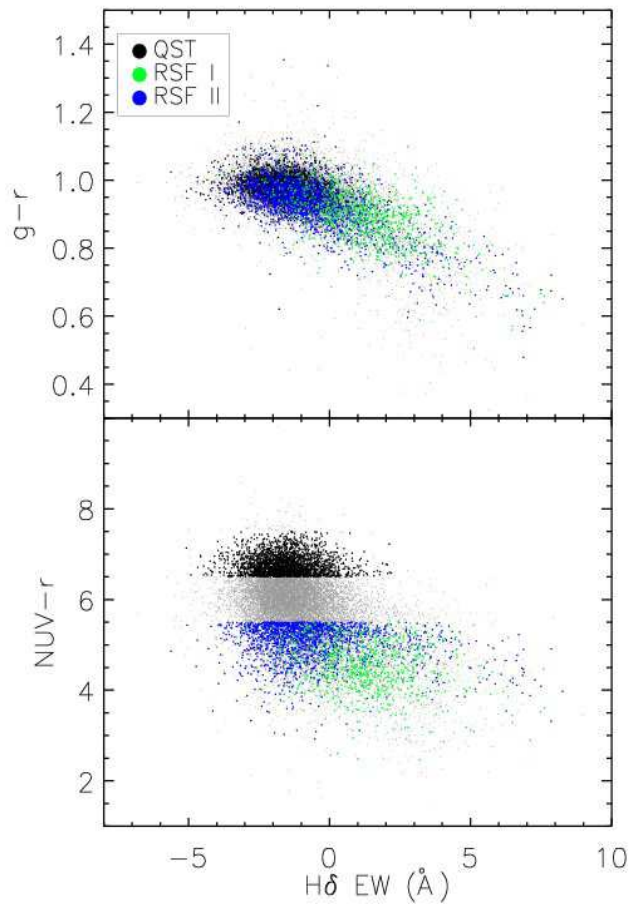}
\caption{The H$\delta$ vs. $g-r$ diagram (upper panel) and the
H$\delta$ vs. $NUV-r$ diagram (lower panel) for QST, RSF I, and RSF II.
Both RSF I and RSF II are indistinguishable from QST in the $g-r$ colour
distribution. Contrary to $g-r$ colour, there is a separation between
RSF I\,/\,II and QST in the $NUV-r$ colour distribution.
A gap between QST and RSF I\,/\,II in $NUV-r$ colour is an artifact of
their different selection criteria.}
\label{fig:grnuvr}
\end{center}
\end{figure}

Figure~\ref{fig:grnuvr} illustrates the power of UV detecting young stellar populations. Most of the
quiescent and UV-excess early-type galaxies are so narrowly distributed in
$g-r$ colour ($\sim$ 0.2 mag) that UV-excess early-type galaxies
cannot be distinguished from the quiescent ones. In contrast,
early-type galaxies with RSF are clearly separated from the quiescent ones in $NUV-r$ colour.
Note that the FUV flux is often affected by the UV upturn
while the NUV flux is less sensitive to the phenomenon \citep{yoon04b,lee05g1,ree07}.
Thus, the NUV flux is a better tracer for RSF activities in early-type galaxies. This is why
$NUV-r$ colour is used as a selection criterion for UV-excess early-type
galaxies. The criterion for UV-excess early-type galaxies is rather conservatively defined as
$NUV-r$ $<$ 5.5 \citep{yi05,kaviraj07b,schawinski07}. The threshold comes from the
$NUV-r$ colours of nearby giant early-type galaxies with the strongest UV upturn but no sign
of RSF activities \citep[e.g][]{burstein88,lee05g1,ree07}.
Therefore, we can not rule out the possibility that this threshold excludes
some of the less massive RSF galaxies, but, a better threshold than this should
await more precise information about the dependence of UV upturn properties
on galaxy mass.

The selected UV-excess early-type galaxies are further divided into the following
two types by using $NUV-r$ colour and another prime RSF indicator,
H$\alpha$ emission line: (a) Ongoing weak star formation galaxies (RSF I) that
are characterized by the presence of H$\alpha$ emission line with
an RSF signature in $NUV-r$ colour, and (b) Post-weak-star-formation or bygone
weak star formation galaxies (RSF II) that show a sign of
RSF in $NUV-r$ colour but no H$\alpha$ emission. The rest of the galaxies are
UV-dead quiescent galaxies that show no evidence of new-born stars, and
are referred to as ``QST'' .

%% Park&Choi scheme
\begin{figure}
\begin{center}
\includegraphics[width=8.5cm]{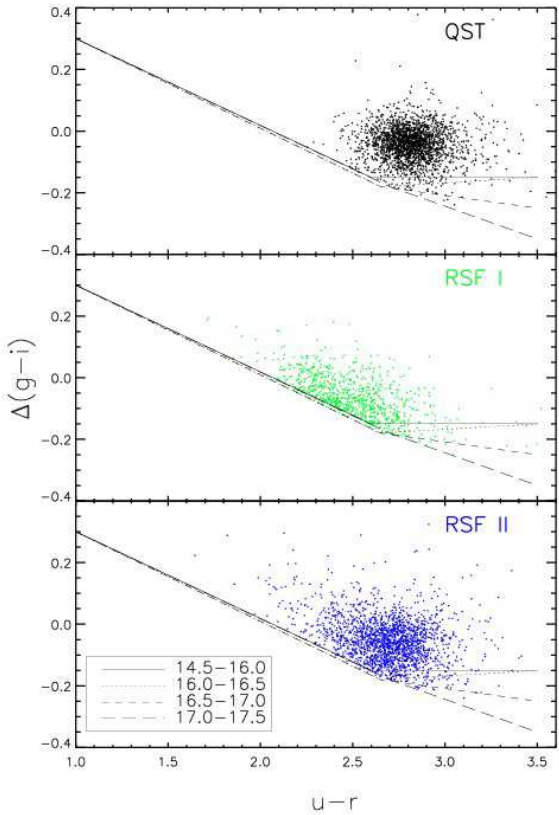}
\caption{Distributions of QST (top), RSF I (middle), and RSF II (bottom) galaxy subsamples in the $u-r$ colour vs. radial $g-i$ colour gradient plane. This is the same as Fig. 1 (lower panel) of \citet{park05}, but with early-type galaxies having GALEX UV information. The exact shape of each line is as a function of galaxy $r_{pet}$ as denoted in the bottom panel.}
\label{fig:scheme}
\end{center}
\end{figure}
The \citet{park05} morphological classification method performs well in general.
However, since the strategy relies on galaxy colours, the UV-excess early-type
galaxies may be selectively missed in this scheme. It is important to specify
how this affect our results. Figure~\ref{fig:scheme} is the same as Fig. 1
(lower panel) of \citet{park05} but with early-type galaxies having
GALEX UV information. It is clear that QST and RSF II galaxies are fairly
secure in this scheme, while about a quarter of RSF I galaxies appear to
be more consistent with the late-type morphology. This agrees with the
result based on visual inspections in Figures~\ref{fig:img_qst},~\ref{fig:img_sf2},
and~\ref{fig:img_ea}. Since our results are based mainly on the QST and RSF II galaxies,
the colour-based classification scheme does not affect our conclusion.

In addition, we select 1,548 early-type galaxies with a strong
H$\alpha$ emission line from the SDSS DR6 with the following criteria:
\begin{itemize}
\item Inverse concentration index, $r_{90}$/$r_{50}$ $<$ 0.33
\item $b/a$ (axis ratio) in the $r$-band $\ge$ 0.6
\item H$\alpha$ emission line EW $\le$ $-$9 \AA
\item S/N in the $r$-band $>$ 10
\item AGN-free on the BPT diagram \citep{baldwin81,kewley01,goto05L}
\item Exclusion of 36 disc bulges by eye inspection
\end{itemize}
We refer to these galaxies as ``SH$\alpha$'' (strong H$\alpha$) and they do
not necessarily have the GALEX UV data.
%
% QST, RSF I\,/\,II, SHa Properties
\begin{table}
 \centering
  \caption{Four different RSF modes in early-type galaxies. Early-type galaxies
  in our sample are classified by using H$\alpha$ emission line and $NUV-r$ colour.}
  \label{tab:ETGpro}
  \begin{tabular}{@{}lccc@{}}
  \hline
  \hline
    RSF mode & H$\alpha$ & $NUV-r$ & Class  \\
             & Emission line  & colour  \\
 \hline
 Quiescent mode  &            No            &             Red          & QST \\
                 & ($\ge$ 0.5 \AA) & (6.5 $\le$ , $\le$ 7.5)    &     \\
 %\hline
 Ongoing weak SF &     Weak                 &           Blue           & RSF I \\
                 & ($-$9 \AA $\le$ , $\le$ $-$3 \AA) &  ($<$ 5.5)         &       \\
 %\hline
 Bygone weak SF    &       No                 &           Blue             & RSF II \\
                 & ($\ge$ 0.5 \AA) & ($<$ 5.5)              &``E+a'' \\
 %\hline
 Ongoing starburst &         Strong         &                          & SH$\alpha$ \\
                 & ($\le$ $-$9 \AA)  &                          &  \\
\hline
\end{tabular}
\end{table}
The selection criteria for the different types of galaxies introduced in this
section are summarized in Table~\ref{tab:ETGpro}.
% End of DATA

%
\begin{figure}
\begin{center}
\includegraphics[width=8.5cm]{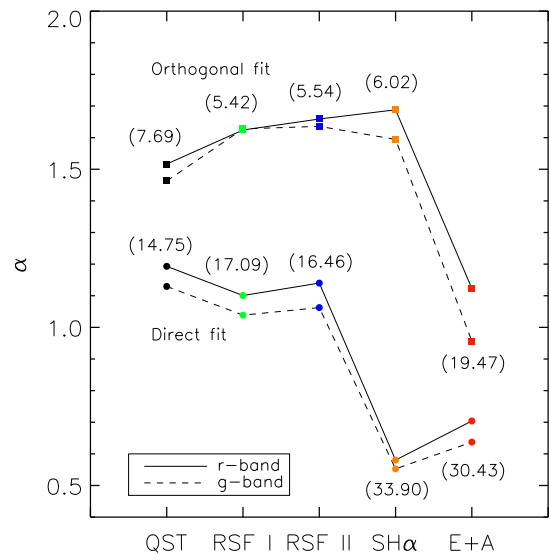}
\caption{The evolution corrected coefficients of all galaxy types obtained from
the direct fitting (circles) and the orthogonal fitting (squares) methods.
Galaxies in different RSF modes have different FP slopes ($\alpha$).
The solid and dashed lines are the coefficients in the $r$-band and the
$g$-band, respectively. The coefficients show wavelength
dependency in all types of galaxies. Note that the FP coefficients vary with
fitting methods. The direct fit gives the steepest slope for the QST FP and
the shallowest for the SH$\alpha$ FP in both bands. The orthogonal fit yields
the steepest slope for the SH$\alpha$ FP in the $r$-band and the RSF II FP
in the $g$-band. The figures in parentheses represent the
dihedral angles between our $g$-band planes and the virial plane in units of degree.}
\label{fig:alpha}
\end{center}
\end{figure}

\section[]{The Fundamental Planes of E+A galaxies and GALEX\,/\,SDSS Early-type galaxies}
\subsection{Construction of the Fundamental Planes}
\label{sec:FP}
%
% Coefficients of FP
\begin{table*}
  \caption{Coefficients of the FPs of the five types of early-type galaxies
  in the $r$-band and the $g$-band. Errors are estimated from the bootstrap method
  (10,000 resamplings). The rms of the direct fit and the rms of the orthogonal fit
  are the scatter around the plane in the log $R_{e}$ direction and in the perpendicular direction,
  respectively. We apply the evolution correction following the evolution parameter
  Q of \citet{bernardi03c}.}
  \label{tab:coeffi}
  \begin{tabular}{@{}lccccccccc@{}}
  \hline
  \hline
           & \multicolumn{5}{c}{$r$-band}          & \multicolumn{4}{c}{$g$-band} \\
  %\hline
                & $\alpha$ & $\beta$ & $\gamma$ & rms & & $\alpha$ & $\beta$ & $\gamma$ & rms \\
  \hline
  \hline
   Direct Fit \\
  %\hline
    QST         & 1.25\,$\pm$\,0.01 & 0.310\,$\pm$\,0.003 & --8.33\,$\pm$\,0.07 & 0.058 && 1.20\,$\pm$\,0.01 & 0.307\,$\pm$\,0.003 & --8.46\,$\pm$\,0.08 & 0.047 \\
    RSF I       & 1.13\,$\pm$\,0.03 & 0.254\,$\pm$\,0.006 & --6.91\,$\pm$\,0.16 & 0.166 && 1.08\,$\pm$\,0.03 & 0.238\,$\pm$\,0.006 & --6.68\,$\pm$\,0.17 & 0.166 \\
    RSF II      & 1.18\,$\pm$\,0.02 & 0.254\,$\pm$\,0.005 & --7.06\,$\pm$\,0.11 & 0.160 && 1.12\,$\pm$\,0.02 & 0.239\,$\pm$\,0.005 & --6.83\,$\pm$\,0.11 & 0.156 \\
    SH$\alpha$  & 0.59\,$\pm$\,0.03 & 0.200\,$\pm$\,0.008 & --4.60\,$\pm$\,0.19 & 0.191 && 0.57\,$\pm$\,0.03 & 0.222\,$\pm$\,0.006 & --5.14\,$\pm$\,0.15 & 0.194 \\
    E+A         & 0.73\,$\pm$\,0.03 & 0.206\,$\pm$\,0.004 & --4.98\,$\pm$\,0.12 & 0.184 && 0.68\,$\pm$\,0.02 & 0.209\,$\pm$\,0.003 & --5.05\,$\pm$\,0.10 & 0.170 \\
  \hline
  Orthogonal Fit  \\
  %\hline
    QST         & 1.61\,$\pm$\,0.02 & 0.306\,$\pm$\,0.004 & --9.09\,$\pm$\,0.08 & 0.053 && 1.60\,$\pm$\,0.02 & 0.301\,$\pm$\,0.004 & --9.23\,$\pm$\,0.09 & 0.055  \\
    RSF I       & 1.72\,$\pm$\,0.05 & 0.286\,$\pm$\,0.007 & --8.82\,$\pm$\,0.22 & 0.072 && 1.75\,$\pm$\,0.05 & 0.266\,$\pm$\,0.007 & --8.71\,$\pm$\,0.23 & 0.078  \\
    RSF II      & 1.75\,$\pm$\,0.03 & 0.270\,$\pm$\,0.005 & --8.62\,$\pm$\,0.13 & 0.075 && 1.75\,$\pm$\,0.03 & 0.249\,$\pm$\,0.005 & --8.43\,$\pm$\,0.14 & 0.080  \\
    SH$\alpha$  & 1.80\,$\pm$\,0.12 & 0.230\,$\pm$\,0.011 & --7.83\,$\pm$\,0.34 & 0.129 && 1.76\,$\pm$\,0.12 & 0.251\,$\pm$\,0.009 & --8.30\,$\pm$\,0.33 & 0.130  \\
    E+A         & 1.20\,$\pm$\,0.04 & 0.222\,$\pm$\,0.005 & --6.35\,$\pm$\,0.14 & 0.108 && 1.05\,$\pm$\,0.04 & 0.217\,$\pm$\,0.003 & --6.06\,$\pm$\,0.12 & 0.104 \\
  %\hline
  \hline
  Evolution-Corrected Direct Fit \\
  %\hline
    QST         & 1.19\,$\pm$\,0.01 & 0.311\,$\pm$\,0.003 & --8.23\,$\pm$\,0.07 & 0.014 && 1.12\,$\pm$\,0.01 & 0.308\,$\pm$\,0.003 & --8.32\,$\pm$\,0.07 & 0.035  \\
    RSF I       & 1.10\,$\pm$\,0.03 & 0.258\,$\pm$\,0.006 & --6.92\,$\pm$\,0.15 & 0.159 && 1.03\,$\pm$\,0.03 & 0.244\,$\pm$\,0.006 & --6.71\,$\pm$\,0.16 & 0.156  \\
    RSF II      & 1.14\,$\pm$\,0.02 & 0.258\,$\pm$\,0.005 & --7.06\,$\pm$\,0.11 & 0.150 && 1.06\,$\pm$\,0.02 & 0.244\,$\pm$\,0.004 & --6.82\,$\pm$\,0.11 & 0.142  \\
    SH$\alpha$  & 0.58\,$\pm$\,0.03 & 0.211\,$\pm$\,0.008 & --4.79\,$\pm$\,0.18 & 0.183 && 0.55\,$\pm$\,0.03 & 0.230\,$\pm$\,0.006 & --5.27\,$\pm$\,0.14 & 0.182  \\
    E+A         & 0.70\,$\pm$\,0.03 & 0.210\,$\pm$\,0.004 & --5.02\,$\pm$\,0.12 & 0.175 && 0.63\,$\pm$\,0.02 & 0.211\,$\pm$\,0.003 & --5.03\,$\pm$\,0.09 & 0.161  \\
  %\hline
  \hline
  Evolution-Corrected Orthogonal Fit \\
  %\hline
    QST         & 1.51\,$\pm$\,0.02 & 0.306\,$\pm$\,0.003 & --8.87\,$\pm$\,0.07 & 0.051 && 1.46\,$\pm$\,0.02 & 0.300\,$\pm$\,0.003 & --8.94\,$\pm$\,0.07 & 0.053  \\
    RSF I       & 1.62\,$\pm$\,0.05 & 0.284\,$\pm$\,0.007 & --8.60\,$\pm$\,0.20 & 0.071 && 1.62\,$\pm$\,0.05 & 0.264\,$\pm$\,0.007 & --8.44\,$\pm$\,0.21 & 0.076  \\
    RSF II      & 1.65\,$\pm$\,0.03 & 0.270\,$\pm$\,0.005 & --8.43\,$\pm$\,0.12 & 0.074 && 1.63\,$\pm$\,0.03 & 0.249\,$\pm$\,0.005 & --8.20\,$\pm$\,0.13 & 0.079  \\
    SH$\alpha$  & 1.68\,$\pm$\,0.12 & 0.232\,$\pm$\,0.011 & --7.63\,$\pm$\,0.31 & 0.127 && 1.59\,$\pm$\,0.11 & 0.251\,$\pm$\,0.008 & --7.95\,$\pm$\,0.30 & 0.128  \\
    E+A         & 1.12\,$\pm$\,0.04 & 0.223\,$\pm$\,0.004 & --6.20\,$\pm$\,0.13 & 0.105 && 0.95\,$\pm$\,0.03 & 0.217\,$\pm$\,0.003 & --5.87\,$\pm$\,0.11 & 0.102  \\
  \hline
\end{tabular}
\end{table*}
The three observed parameters for the FP analysis are obtained as follows.
We set the effective radius to $r_{e}$ $\equiv$ $(b/a)^{0.5}$$r_{dev}$,
where $b/a$ is the ratio of the minor and major axes and $r_{dev}$ is an
effective radius from the de Vaucouleurs fit \citep{devau48}. We do this
in order to convert an elliptical aperture to an effective circular radius
\citep{bernardi03a}.
The effective mean surface brightness within $r_{e}$ is given by
$\mu$ = $m_{dev}$ + 2.5 log(2$\pi$$r_{e}^{2}$) $-$ K($z$) $-$ 10 log($z$+1),
where $m_{dev}$ is the apparent magnitude obtained by the de Vaucouleurs' $r^{1/4}$
fit \citep{devau48} and $z$ is the redshift. We take into account the K-correction
as described by \citet{blanton03} and the cosmological (1+$z$)$^{4}$ dimming effect
\citep[e.g.][]{tolman30}.
We correct the velocity dispersion for the aperture effect of the SDSS using
$\sigma_{cor}$ = $(\frac{r_{fiber}}{r_{e}/8})^{0.04}\sigma_{est}$,
where $r_{fiber}$ is 1.5 arcsec and $r_{e}$ is in arcsec \citep{jorgensen95}.

We determine the FP coefficients ($\alpha$, $\beta$, $\gamma$) of each type of
galaxies (QST, RSF I\,/\,II, SH$\alpha$, and E+A) by using the two schemes;
the least square direct fitting method and the least square orthogonal fitting method.
By performing the robust fit, we reduce the effect of outliers.
Direct fitting yields the coefficients that minimize the thickness of the plane
in the log $R_{e}$ direction, while the orthogonal fitting algorithm
minimizes the summation of residuals perpendicular to the plane.
The plane coefficients are obtained from simple algebraic calculations.
For the orthogonal fitting, we use a following covariance matrix
%
% Covariance Matrix
\begin{displaymath}
\mathbf{\textsf{\textbf{C}}} =
\left( \begin{array}{ccc}
\sigma_{V}^2      & \sigma_{V\mu}^2    & \sigma_{VR}^2  \\
\sigma_{V\mu}^2   & \sigma_{\mu}^2     & \sigma_{R\mu}^2  \\
\sigma_{VR}^2     & \sigma_{R\mu}^2    & \sigma_{R}^2
\end{array} \right)
\end{displaymath}
for which the eigenvector corresponding to the smallest eigenvalue is the normal vector
of the plane we are seeking. The smallest eigenvalue represents the rms scatter
of the plane \citep[e.g.][]{saglia01,bernardi03c}. For the direct fitting method,
on the other hand, we use the same algebra as \citet{bernardi03c}.
The uncertainties in the coefficients are estimated via 10,000 bootstrap
resamplings of the data. The best-fit coefficients and their errors are presented in
Table~\ref{tab:coeffi}.

For a sanity check, we make a combined subset of QST and week-SF (i.e. RSF I and RSF II)
galaxies and compare the FP coefficients to those of \citet{bernardi03a}. We assume that
by putting SH$\alpha$ galaxies aside the subset mimics their sample.
The derived coefficients from both the direct fit and the orthogonal fit are about 5 $\sim$ 10 per
cent larger than those of \citet{bernardi03c}. \cite{bernardi03b} showed that the corrections
for the evolution and selection effects tend to reduce the coefficient $\alpha$.
They determined the evolution parameter Q using the maximum
likelihood method in the $g$-band (0.85) and in the $r$-band (1.15).
After taking weak passive luminosity evolution effect into account,
we find the coefficients for the combined sample in very good agreement
with \citet{bernardi03c}. For example, the evolution corrected coefficients
$\alpha$ of our combined sample are 1.07 for the direct fit
and 1.50 for the orthogonal fit in the $g$-band. The Q-correction has been applied to all classes
of our sample (QST, RSF I\,/\,II, SH$\alpha$, and E+A). The slopes of these
subclasses become shallower and consistent with the coefficients suggested by
\citet{bernardi03c} within the coefficient measurement errors.
The evolution corrected coefficients are listed in Table~\ref{tab:coeffi}.
Note that we do not apply the correction for the selection effect which
is no more straightforward due to complications arising from the cross matching
between two datasets (SDSS and GALEX). Nevertheless, the bias in $\alpha$ due to
this effect is relatively insignificant compared to the evolution effect
\citep{bernardi03c} and hence does not affect our results.

Figure~\ref{fig:alpha} shows the evolution corrected coefficients $\alpha$
of all galaxy types obtained from the direct fit (circles) and the orthogonal
fit (squares). The well-known wavelength dependency of the FP coefficients
\citep[e.g.][]{pahre98,scodeggio98} is weak \citep[e.g][]{bernardi03c,colless01}
but apparent. %The FP coefficients vary with fitting methods.
It has been pointed out that the choice of data fitting method
affects the FP coefficients \citep[e.g.][]{saglia01,bernardi03c,donofrio08}.
The systematic difference among various fitting methods is most likely
caused by the scatter of the FP \citep[e.g.][]{barbera00}.
Given that there is no standard method for deriving the coefficients, it is
important to choose the method that describes the given data most properly.
The coefficients $\alpha$ are especially important values when they are used as
indicators of the FP tilt. To test which fitting methods work as more reliable
and robuster indicators of the FP tilt, we examine a dihedral angle between our plane
and the virial plane. The dihedral angles in degree are denoted in parentheses
(Fig.~\ref{fig:alpha}). While the coefficients $\alpha$ from the orthogonal fit
do not show a systematic tendency, those from the direct fit behave
well as a function of the dihedral angles, in the sense that
the coefficients $\alpha$ decrease as the dihedral angles increase.
For the orthogonal fit, the rank of $\alpha$ among different galaxy types
can even change depending on the band in use. For instance, the orthogonal fit gives the steepest slope
(i.e. the greatest $\alpha$) for SH$\alpha$ in the $r$-band and for RSF II in
the $g$-band, respectively. It is also worth noting that while the orthogonal
fit has the advantage of treating the FP parameters symmetrically, the method
yields coefficients with larger uncertainties, especially for SH$\alpha$.
Taken together, the coefficients $\alpha$ from the direct fit more
successfully describe the relative features of the FP slopes among different
galaxy types. Hence, we use the coefficients $\alpha$ based on the direct
fitting method in the following analysis.
%
% FP of E+A with 4 types of ETGs
\begin{figure*}
\begin{center}
\includegraphics[height=13cm]{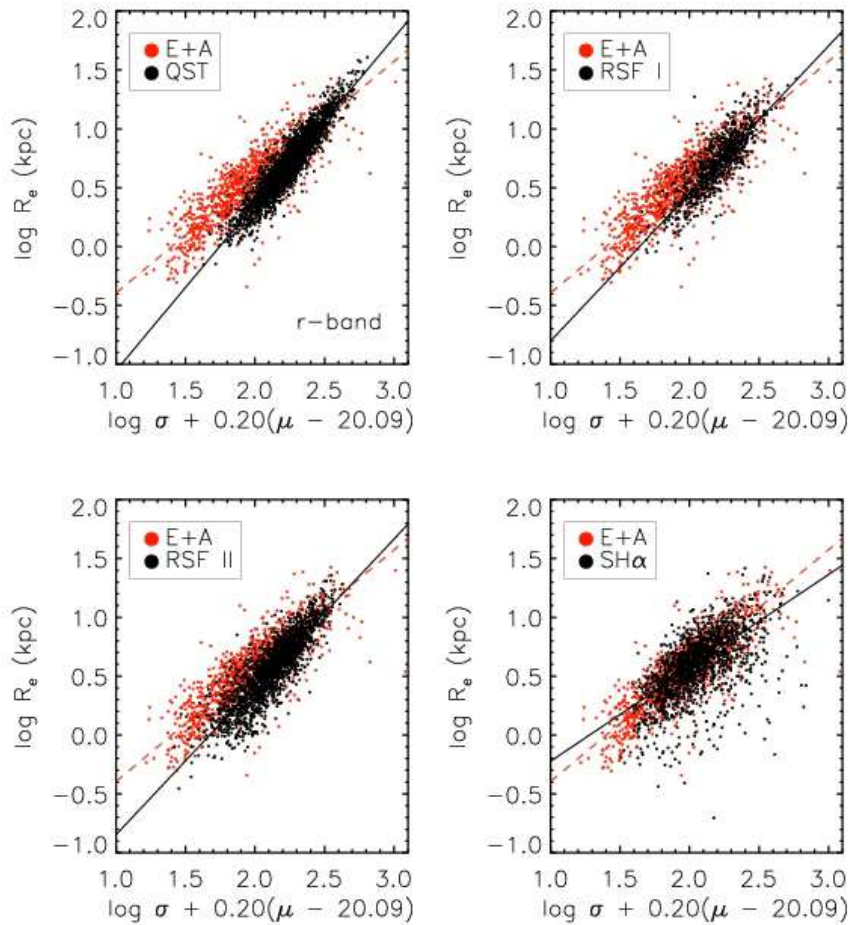} %width=13.5cm,
\vspace*{2mm}
\caption{Edge-on view of the $r$-band FP for five different types of early-type galaxies.
Red circles are E+A (every panel) and black circles are QST (upper-left panel), RSR I (upper-right),
RSF II (lower-left), and SH$\alpha$ (lower-right), respectively. Lines represent the best fit of each
sample in the edge-on projection of the virial plane. The distribution of E+A is most analogous
to that of strong H$\alpha$ galaxies. RSF II also exhibits relatively strong similarities to E+A.}
\label{fig:ea-r}
\end{center}
\end{figure*}

\subsection{Comparison of the Fundamental Planes of E+As and Early-type Galaxies in Different RSF Modes}
\label{sec:EA}
\begin{figure*}
\begin{center}
\includegraphics[height=13cm]{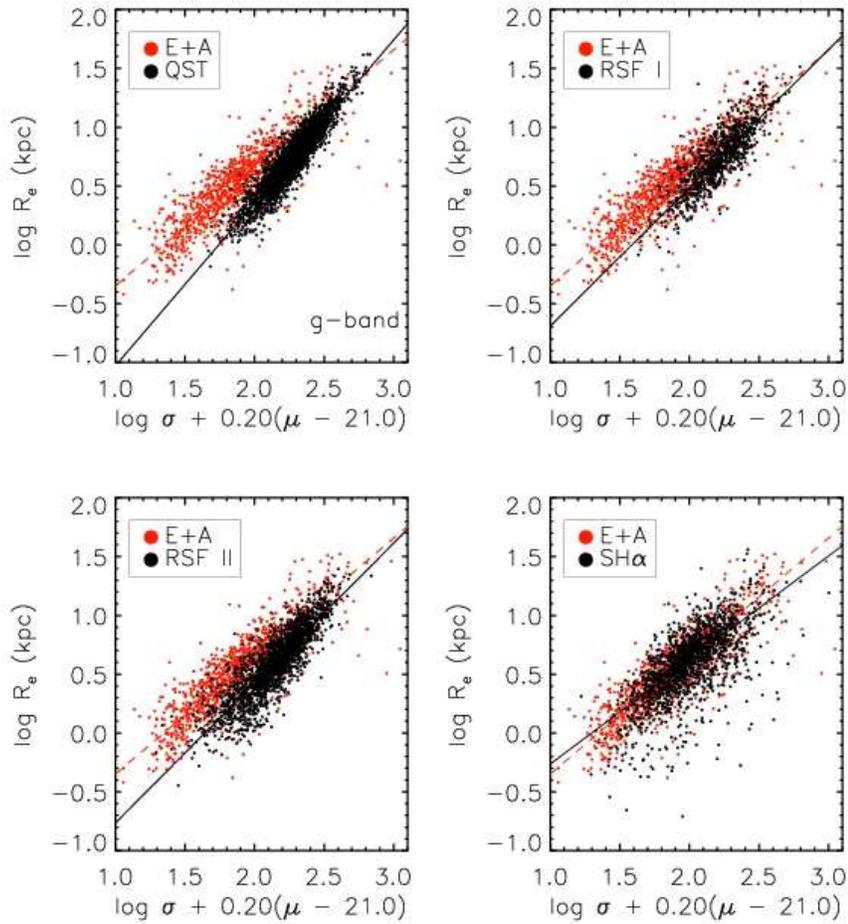} %width=13.5cm,
\vspace*{2mm}
\caption{Same as Fig.~\ref{fig:ea-r}, but in the $g$-band. Since the $g$-band
is more sensitive to the young stellar population than the $r$-band,
the separation among different types becomes more obvious.}
\label{fig:ea-g}
\end{center}
\end{figure*}
\citet{yang08} measured the coefficients of the FP of 16 E+As in the Gunn $r$ and
the Johnson $B$ using the orthogonal fitting method. Their best-fit FP has
$\alpha$ = 1.13\,$\pm$\,0.10 and $\beta$ = 0.24\,$\pm$\,0.07 for the Gunn $r$, and
$\alpha$ = 1.09\,$\pm$\,0.08 and $\beta$ = 0.23\,$\pm$\,0.06 for the Johnson $B$.
The largest E+A sample ($\sim$ 60 times larger than that used in Yang et al. 2008)
allows us to estimate the E+A FP with higher accuracy.
Our E+A FP based on the orthogonal fitting method is given by
\begin{eqnarray}
\log R_{e} &=& 1.20(1.12)\,\log \sigma \, + \,0.222(0.223)\,\mu \, - \,6.35(6.20)\,,
\nonumber \\
\log R_{e} &=& 1.05(0.95)\,\log \sigma \, + \,0.217(0.217)\,\mu \, - \,6.06(5.87)
\nonumber
\end{eqnarray}
in the $r$- band and $g$-band, respectively. The evolution corrected coefficients
appear in parentheses.
When corrected for the evolution effect, $\alpha$ becomes 1.12 in the $r$-band
and 0.95 in the $g$-band, while $\beta$ remains nearly constant in both bands.
Although the bandpass used in \citet{yang08} differs slightly from the SDSS
$r$-band and $g$-band, our E+A FP obtained from the orthogonal fit agrees
well with that of \citet{yang08} within the uncertainties regardless of whether
the evolution effect is corrected or not. The E+A FP calculated by the direct
fitting method is
\begin{eqnarray}
\log R_{e} &=& 0.73(0.70)\,\log \sigma \, + \,0.206(0.210)\,\mu \, - \,4.98(5.02)\,,
\nonumber \\
\log R_{e} &=& 0.68(0.63)\,\log \sigma \, + \,0.209(0.211)\,\mu \, - \,5.05(5.03)
\nonumber
\end{eqnarray}
in the $r$-band and the $g$-band, respectively.

Figures~\ref{fig:ea-r} and~\ref{fig:ea-g} show the distribution of E+A
on the edge-on view of the virial plane, as compared to four other types
of galaxies; QST, RSF I, RSF II, and SH$\alpha$. The differences between
E+A and the other three types of galaxies are more clearly seen in the
$g$-band (see Fig.~\ref{fig:ea-g}), which is more sensitive to the
presence of young stars than the $r$-band.
The comparison shows that E+A lies on a plane that is different from
those of QST, RSF I, and RSF II in terms of slope, position, and scatter.
As mentioned in Section~\ref{sec:FP}, the coefficients $\alpha$ obtained
from the direct fitting method are used when comparing the FP slopes.
It is evident that E+A is absent in the upper right part and more extended
toward the lower left part of the edge-on projection.
Among the four types of galaxies, the QST FP is the most distinct from the E+A FP,
both in slope and in scatter. The FP slope of E+A corresponds to $\sim$ 60 per cent of the
QST slope in both the $r$-band and the $g$-band, and the rms scatter (i.e. the thickness) of the
E+A FP is nearly three times greater than that of the QST FP. In terms of position, the E+A FP
extends more toward the lower left part as compared to the QST FP.
Comparing RSF I to E+A, we find a similar slope, but a significantly different
position. RSF I is distributed over a narrow range, whereas E+A occupies a relatively
long and broad region on the edge-on projection of the virial plane. On the contrary,
RSF II is similar to E+A in both the slope and position. The SH$\alpha$ FP shows
a stronger similarity to that of E+A with slightly larger (1.15 times) scatter.
The scatter around the SH$\alpha$ FP is the largest, and the scatter of the E+A,
RSF I\,/\,II, and QST follow it in the order.
%
% compare the distribution of parameters of E+A with those of QST
\begin{figure*}
\begin{center}
\includegraphics[width=10cm,height=10cm]{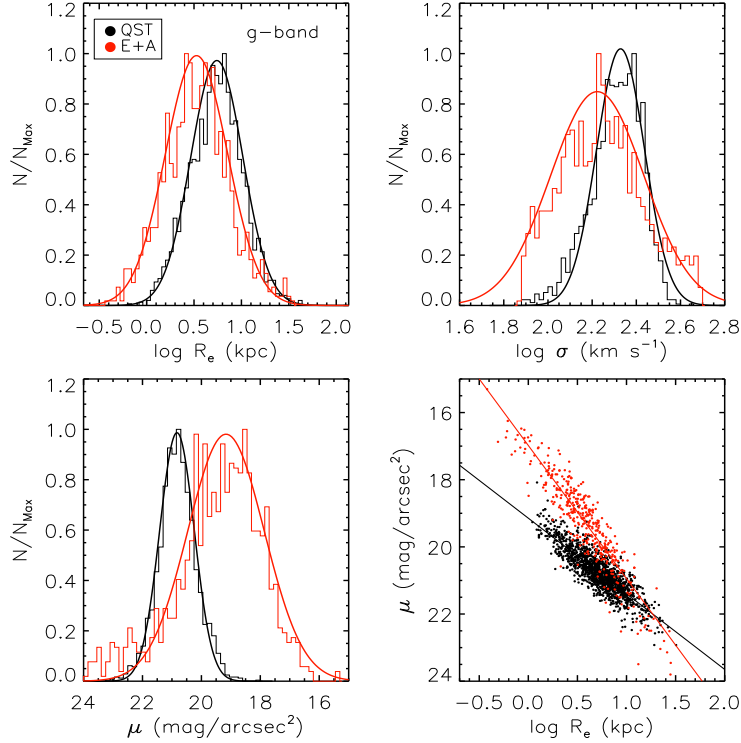}
\vspace*{2mm}
\caption{The normalized distributions of log $R_{e}$, log $\sigma$, and $\mu$
of E+A and QST in the $g$-band. The black and red histograms present the distributions
of the FP parameters of QST and E+A, respectively. Solid curves are fitted
Gaussian functions. In all parameters, E+A shows a larger standard deviation and a smaller mean
than QST. The lower right panel shows the Kormendy relation of E+A and QST with 2.2
$\le$ log $\sigma$ $<$ 2.4. The Kormendy relation of E+A
is steeper than that of QST, leading to the tilt of the E+A FP.}
\label{fig:tilt}
\end{center}
\end{figure*}

\subsection{Origin of the tilt and shift of the E+A Fundamental Plane}
\label{sec:tilt}
In Figure~\ref{fig:tilt}, we compare the distributions of three FP
parameters of QST (black) and E+A (red) in order to investigate the physical
cause of the shift and tilt of the E+A FP.
Each distribution is normalized by its maximum value. The solid curves
are the Gaussian fits to the histograms, the standard deviation and the mean
of which are summarized in Table~\ref{tab:gauss}.
Compared to QST, E+A has smaller means and larger standard deviations
in all parameters of interest.
The smaller effective radius and more luminous surface brightness (i.e. smaller
$\mu$ value) of E+A are likely due to centrally concentrated star formation
\citep{caldwell99}.
A Starburst in the galaxy centre increases the central luminosity, leading to both
a decreased effective radius and increased mean surface brightness within
the effective radius (left panels of Fig.~\ref{fig:tilt}). As a consequence,
E+A migrates toward the lower left corner of the virial plane projection,
as compared to QST (see Fig.~\ref{fig:ea-r} and Fig.~\ref{fig:ea-g}).
The smaller velocity dispersion of E+A is also involved in the shift of the E+A FP.
According to the downsizing scenario, less massive galaxies have experienced
more active star formation in the recent epoch, and so one would
expect E+A to be less massive and thus have a smaller velocity dispersion than QST.

\begin{figure*}
\begin{center}
\includegraphics[height=7cm]{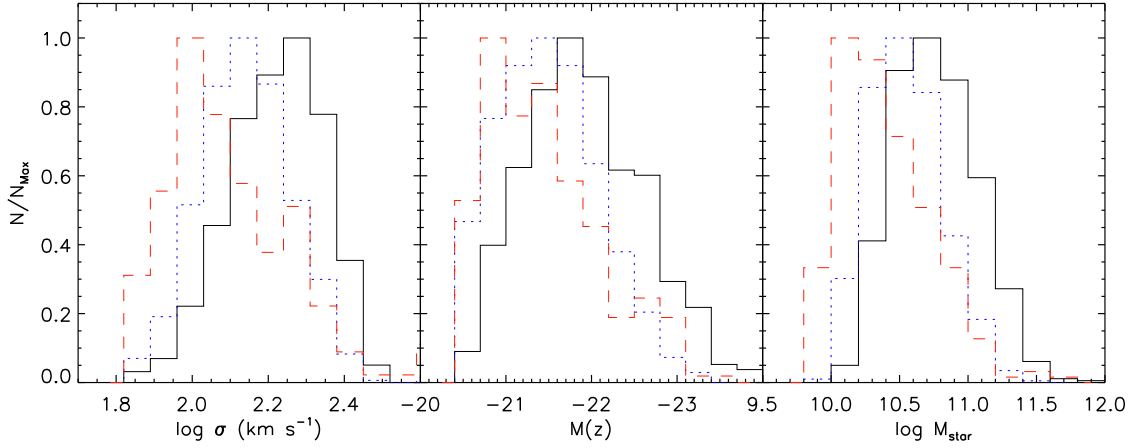} %width=16cm,
\caption{Distributions of velocity dispersion (left panel),
z-band absolute mag (middle panel), and stellar mass (right panel)
for volume-limited early-type subsamples.
The volume-limited sample is constructed based on the following criteria;
0.025 $\le$ $z$ $\le$ 0.09 and M$_{z}$ $\le$ --20.5.
Their stellar masses are obtained from the Garching SDSS catalog
(http://www.mpa-garching.mpg.de/SDSS/DR4/).
The black solid is QST, blue dotted line is RSF II, and red dashed line is E+A.}
\label{fig:mass}
\end{center}
\end{figure*}

% mean and std of fitted gaussian function
\begin{table}
 \centering
  \caption{The means and the standard deviations of the distributions
  for log $R_{e}$, log $\sigma$, $\mu$, and log $\sigma$ + 0.2$\mu$ of all types
  in the $g$-band.}
  \label{tab:gauss}
  \begin{tabular}{@{}lcccc@{}}
  \hline
  \hline
     & {log $R_{e}$} & {log $\sigma$} & {$\mu$} & {log $\sigma$ + 0.2$\mu$}  \\
     & mean~~~  std & mean~~~  std & mean~~~  std & mean~~~  std \\
 \hline
 QST        &  0.740~~~0.274   & 2.328~~~0.109  & 20.828~~~0.589 & 6.480~~~0.182  \\
 RSF I      &  0.743~~~0.248   & 2.187~~~0.148  & 21.153~~~0.758 & 6.417~~~0.182  \\
 RSF II     &  0.608~~~0.296   & 2.219~~~0.147  & 20.756~~~0.774 & 6.365~~~0.206  \\
 SH$\alpha$ &  0.614~~~0.262   & 2.125~~~0.169  & 20.138~~~0.806 & 6.182~~~0.228  \\
 E+A        &  0.526~~~0.328   & 2.223~~~0.211  & 18.922~~~1.269 & 6.067~~~0.316  \\
 \hline
 \end{tabular}
\end{table}
In order to test whether the smaller velocity dispersion of E+A is an
intrinsic property, we construct a volume-limited sample of 0.025
$\le$ $z$ $\le$ 0.09 and M$_{z}$ $\le$ --20.5.
The criteria include galaxies with a wide range of luminosities and a maximum number
of E+As in the sample.
Figure~\ref{fig:mass} (left panel) shows that the volume-limited sample exhibits the same trends as seen in the entire E+A sample, namely, a smaller mean and lager standard deviation in the log $\sigma$ distribution. The mean values of log $\sigma$ are 2.228
for QST and 2.063 for E+A, with standard deviations of 0.135 and
0.192, respectively. One might argue that the velocity dispersion measurement of E+A galaxies may not be a reliable tracer of their masses.
The central regions of E+As will contain young stellar populations, and ordered motion in these central regions could underestimate the velocity dispersions.
Figure~\ref{fig:mass} (middle and right panels), however,
show that the distributions of $M_Z$ and stellar mass are consistent with
that of the velocity dispersion of galaxies,
confirming that the small log $\sigma$ is part of the nature of E+A.
Detailed mass functions of E+As are being investigated by Inami et al. (in preparation).
% Mass
%
% sigma bin mean R, m
\begin{figure*}
\begin{center}
\includegraphics[height=12cm]{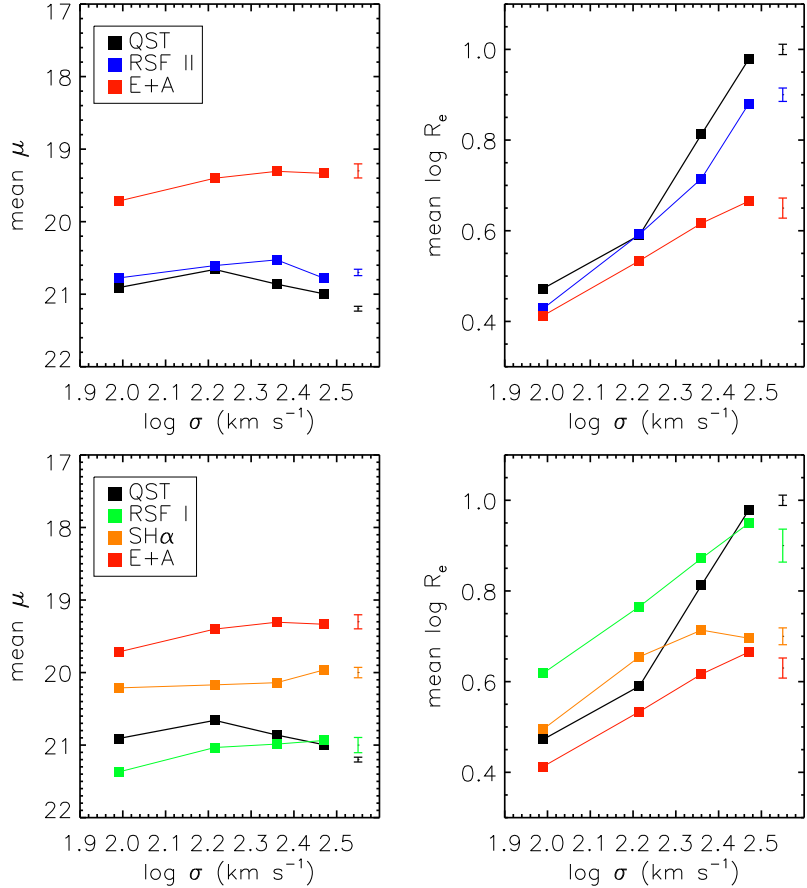} %width=10cm,
\vspace*{2mm}
\caption{The mean log $R_{e}$ and the mean $\mu$ of the five types of galaxies in each log $\sigma$ bin.
We divide $\sigma$, 70 km~s$^{-1}$ to 330 km~s$^{-1}$, into four bins with the bin size of
65 km~s$^{-1}$: 70\,$\sim$\,135, 135\,$\sim$\,200, 200\,$\sim$\,265,
and 265\,$\sim$\,330 km~s$^{-1}$. The corresponding logarithm values are approximately 1.85, 2.13, 2.30, 2.43,
and 2.52. The upper two panels compare QST with RSF II and E+A. The lower two
panels compare QST with RSF I, SH$\alpha$ and E+A. The properties of RSF II are significantly analogous
to those of E+A in both comparisons.}
\label{fig:sigbin}
\end{center}
\end{figure*}

The lower right panel of Figure~\ref{fig:tilt} shows the Kormendy relation
\citep{kormendy77}, a projection of the FP. Black circles and red circles denote QST and
E+A, respectively, of the same log $\sigma$ range (2.2 $\le$ log $\sigma$ $\le$ 2.4).
Solid lines are the best fits of the sample. E+A follows its own
scaling relation. The tendency towards a smaller $r_{e}$ and the more luminous
$\mu$ are apparent: the E+A slope is about 2 times steeper than that of QST,
which we believe is responsible for the tilt of the E+A FP.
In short, both the central concentration of starbursts in E+A and its smaller
velocity dispersion play an important role in the shift and the tilt of the E+A FP.

We have compared the E+A FP with the QST, RSF I/ II, and SH$\alpha$ FPs
in terms of slope, position, and scatter. We wonder whether the four types
of early-type galaxies and E+A are independent or instead associated in some ways.
The origin of the tilt and shift of the E+A FP is also explored by analyzing
its structural and dynamical parameters. When investigating the relationship
among them, the geometry of SF within E+As (i.e. the degree of central concentration)
is believed to be an important key.
We discuss the possible evolutionary connections among
the different types of galaxies in the following section.

\section[]{Discussion}
\subsection{Possible Connection between E+A and UV-excess Galaxies}
\label{sec:evol}
The FP analysis can provide important clues to the physical links among the five
different types of galaxies. We have found that there are systematic
differences in position, slopes and scatter among them (see Section~\ref{sec:EA}).
We have shown that the coefficients increased systematically from SH$\alpha$ (0.55),
to E+A (0.63), to RSF I (1.03) \& II (1.06), and finally to QST (1.12) in the $g$-band.
The systematic change may imply a relationship among the different types of galaxies.
Our main goal here is to examine whether there are any evolutionary paths from
SH$\alpha$ to RSF I, RSF II, or QST via E+A.
% evolutionary connection
\begin{figure}
\begin{center}
\includegraphics[width=8cm,height=8cm]{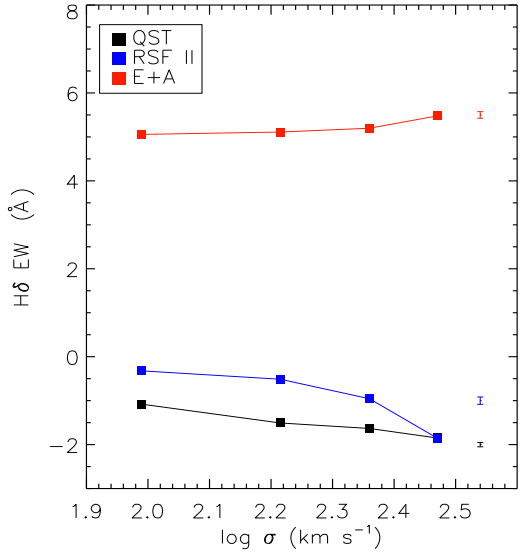}
\caption{Mean H$\delta$ EW of QST, RSF II, and E+A in each log $\sigma$ bin.
The range of log $\sigma$ and the bin size are the same as Fig.~\ref{fig:sigbin}.
Note that RSF II has about 1 $\AA$ larger H$\delta$ EW than QST.}
\label{fig:hd}
\end{center}
\end{figure}
%
%% Figure 12
\begin{figure*}
\begin{center}
\includegraphics[width=15cm,height=15cm]{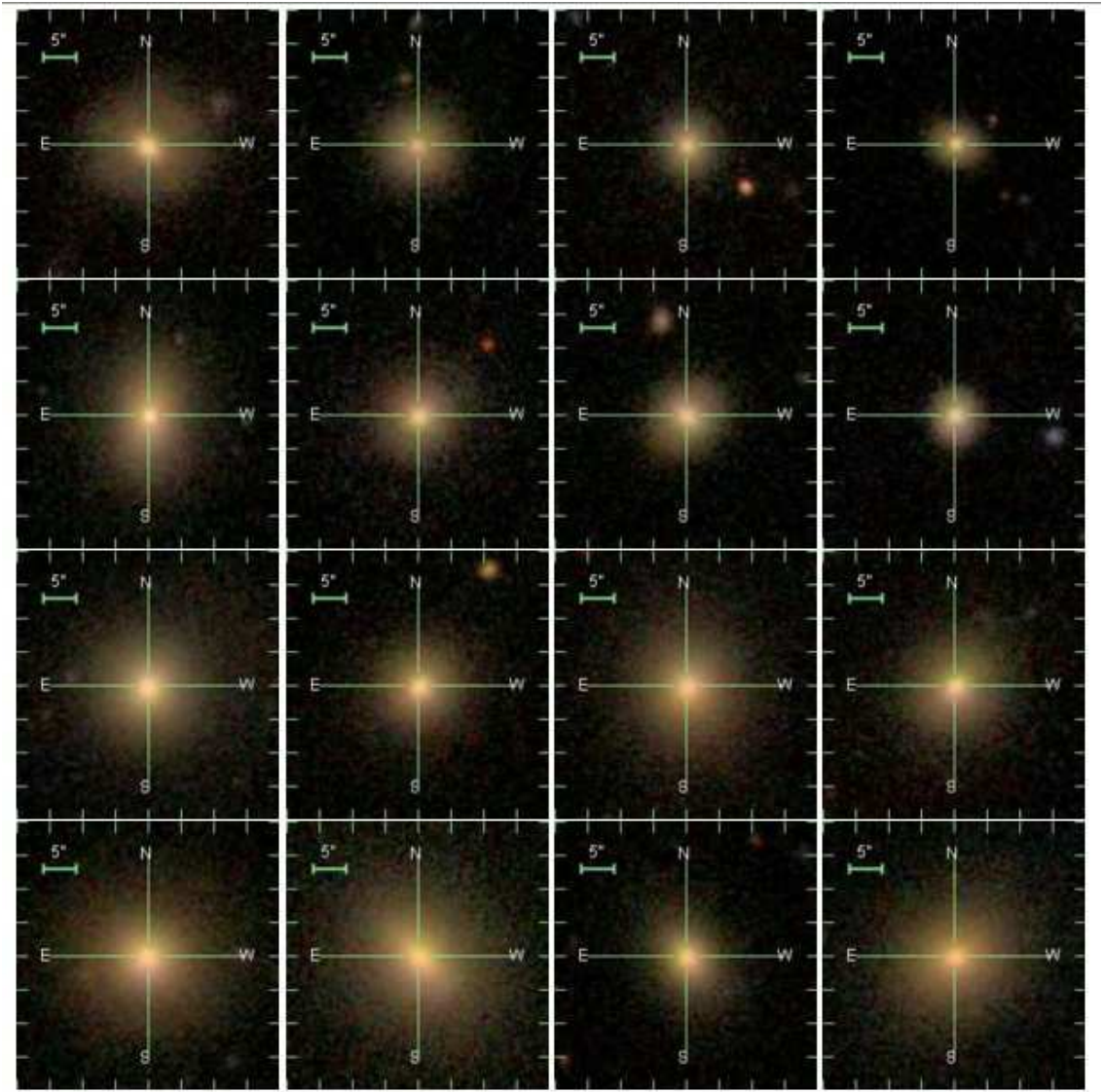}
\caption{The $g$,$r$,$i$ composite SDSS images for QST galaxies in four different log $\sigma$ bins.
The images are sorted by the velocity dispersion:
the 1st row (70\,$\sim$\,135 km~s$^{-1}$),
the 2nd row (135\,$\sim$\,200 km~s$^{-1}$),
the 3rd row (200\,$\sim$\,265 km~s$^{-1}$),
and the last row (265\,$\sim$\,330 km~s$^{-1}$).
Selected are galaxies with redshift ranging between 0.05 and 0.10.
In each row, the 1st (last) column has lowest (highest) redshift.}
\label{fig:img_qst}
\end{center}
\end{figure*}
%
%% Figure 13
\begin{figure*}
\begin{center}
\includegraphics[width=15cm,height=15cm]{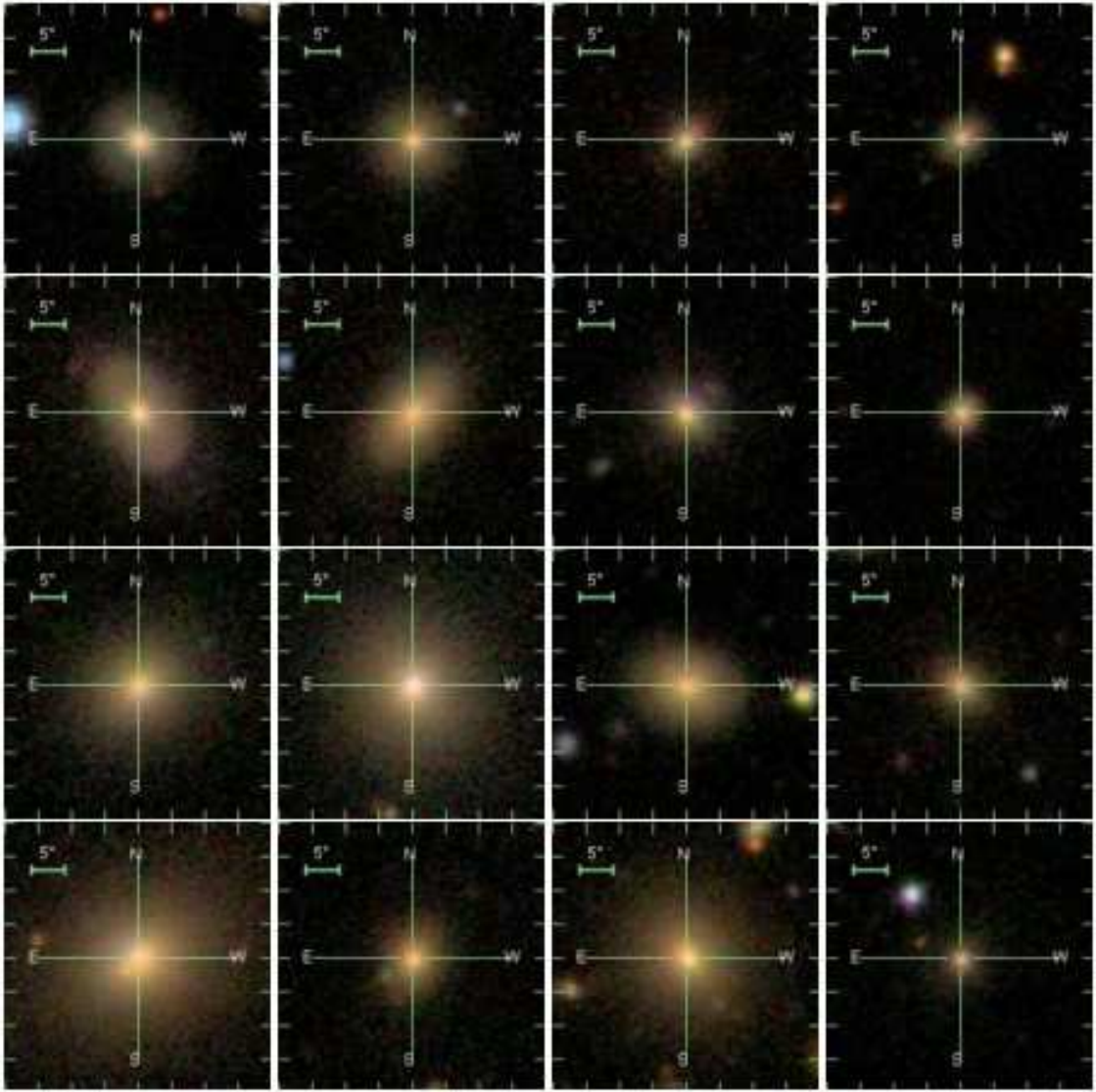}
\caption{The same as Fig. 12, but for RSF II galaxies}
\label{fig:img_sf2}
\end{center}
\end{figure*}
%
%% Figure 14
\begin{figure*}
\begin{center}
\includegraphics[width=15cm,height=15cm]{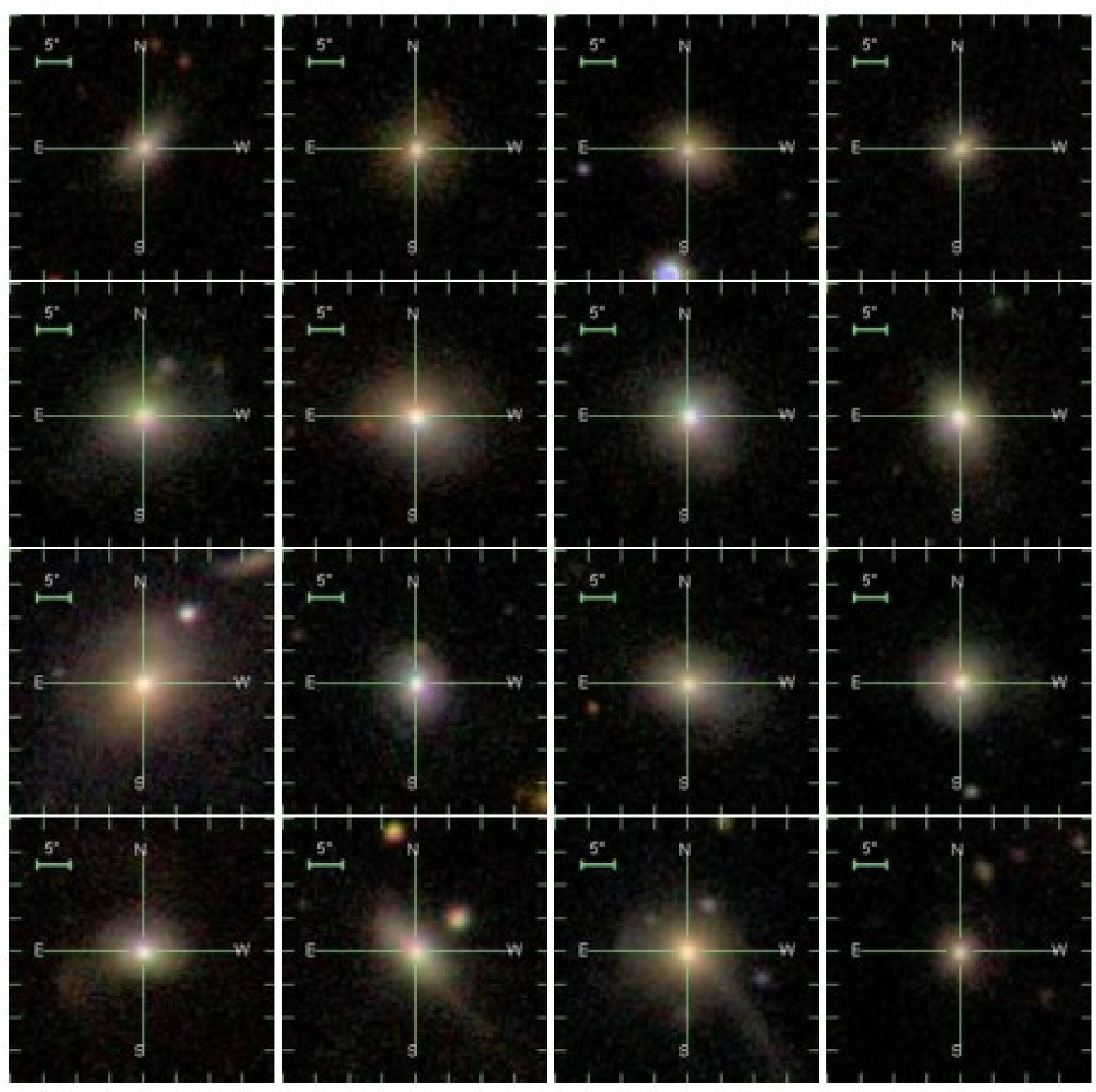}
\caption{The same as Fig. 12, but for E+A galaxies}
\label{fig:img_ea}
\end{center}
\end{figure*}

We explore the distribution of the FP parameters (see Table~\ref{tab:gauss}).
The largest mean effective radius found in RSF I, despite its relatively small mean
velocity dispersion, suggests that weak ongoing star formation in RSF I is not centrally
concentrated, but rather widely spread across the galaxy. The mean surface brightness
of RSF I is fainter than that of QST, which also supports the possibility. RSF II, however,
has a smaller effective radius and more luminous surface brightness than QST, which may
lead the RSF II FP to be analogous to that of E+A. Contrary to our expectation that
SH$\alpha$ is the most similar to E+A in structural and dynamical parameters,
SH$\alpha$ has a significantly larger mean effective radius and less luminous mean surface brightness
than E+A. These results indicate that the ongoing starburst in SH$\alpha$ is not as
highly concentrated as in E+A ($\sim$ 75 per cent of E+As show a positive colour gradient
according to Yamauchi \& Goto 2005), and so the two may have significantly different
RSF geometries. A sudden quench of RSF in SH$\alpha$
would result in E+A with negative or flat colour gradients, rather than E+A with
blue cores or positive colour gradients.

Figure~\ref{fig:sigbin} compares the mean surface brightness and the mean effective
radius of each galaxy type in the same log $\sigma$ bins.
We break $\sigma$, 70 km~s$^{-1}$ to 330 km~s$^{-1}$, into four bins of size 65
km~s$^{-1}$: 70\,$\sim$135, 135\,$\sim$\,200,
200\,$\sim$\,265, and 265\,$\sim$\,330 km~s$^{-1}$. They correspond
approximately to log $\sigma$ $=$ 1.85, 2.13, 2.30, 2.43, and 2.52.
The bin size of 65 km~s$^{-1}$ is determined in order to generate a fair number of
log $\sigma$ bins while retaining statistically meaningful number of galaxies in each bin.
For clarity, we plot QST, RSF II, and E+A in the upper panels and QST, RSF I,
SH$\alpha$, and E+A in the lower panels. The short vertical error bars denote the estimated
uncertainties based on the bootstrap method.
In left panels in Figure~\ref{fig:sigbin}, we compare the mean surface brightness.
RSF II, E+A, and SH$\alpha$ are more luminous than QST. RSF I, however, shows less
luminous mean surface brightness than QST in the first three bins.
The same trend is observed for the mean surface brightness of RSF II and E+A:
an increase in the first three bins followed by a decrease in the last bin.
Unlike RSF II and E+A, the mean surface brightness of SH$\alpha$ remains
roughly constant with log $\sigma$ in the first three bins and increases in the
last bin. The right panels in Fig.~\ref{fig:sigbin} show that
RSF II and E+A have a smaller mean effective radius than QST in all bins,
whereas the mean effective radius of RSF I and even SH$\alpha$ are larger than
that of QST in some bins.

We found no evidence for connection of RSF I or SH$\alpha$ to E+A in this analysis.
It is now clear that even though SH$\alpha$ is found in a very similar
region to the E+A FP, there is no evidence supporting their connection.
Although both RSF I and SH$\alpha$ show signs of ongoing star formation, it is hard to link
them without exact information regarding star formation histories.
The residual star formation in the outskirt of a galaxy can persist
after the extinction of central starburst (due to the gas exhaustion and/or
the AGN feedback). In this process, SH$\alpha$ could turn into RSF I. Or,
it might be that RSF I galaxies are ones that somehow undergo star formation
across the entire galaxy. In this case, RSF I may or may not be related to SH$\alpha$.
In order to understand what causes the RSF I population, we plan to investigate SH$\alpha$
and ultra-luminous infrared galaxies (ULIRGs) in the forthcoming paper
(Bae et al., in preparation).

RSF II displays similar trends to E+A in both mean surface brightness
and mean effective radius, and the two structural parameters of RSF II
fall between those of E+A and QST. These results indicate
that the star formation geometry of RSF II is also concentrated in
the central region as often observed in E+A, even if the influence of the young
stellar population is weaker compared to E+A.
A detailed investigation of the physical geometry of RSF activity
and its impact on the Scaling Relations will be presented in
Choi et al. (in preparation).

Comparisons of the FPs and their parameters for the five types of galaxies
suggest that RSF II may have a more intimate link to E+A than other types.
The Balmer absorption strengths can reveal a possible
transition between E+A and QST via RSF II.
Figure~\ref{fig:hd} presents the mean H$\delta$ EW at each log $\sigma$ bin
for RSF II, QST, and E+A. The mean uncertainty of each type
(vertical error bars), estimated by the bootstrap method, is as small as the symbol
size. As inferred from Figure~\ref{fig:grnuvr}, H$\delta$ EW of RSF II
is in between those of QST and E+A, supporting possible links between RSF II and E+A.
Independently, another useful sanity check on their association comes from
visual inspection of the E+A galaxies. If there is a connection
between E+A and QST classes via the RSF II category,
then there should be morphological similarities which might not
be picked up by the FP parameters.
Figures~\ref{fig:img_qst},~\ref{fig:img_sf2}, and~\ref{fig:img_ea} compare the SDSS images of each class
in four different sigma bins. It is apparent from the figure that the E+A galaxies do look
bulge-dominated and show similarities to QST and RSF II classes.
This is also consistent with RSF II being linked to E+A.

%
% Hd vs. g-r / NUV-r with BC03 Figure
\begin{figure*}
\begin{center}
\includegraphics[width=11cm]{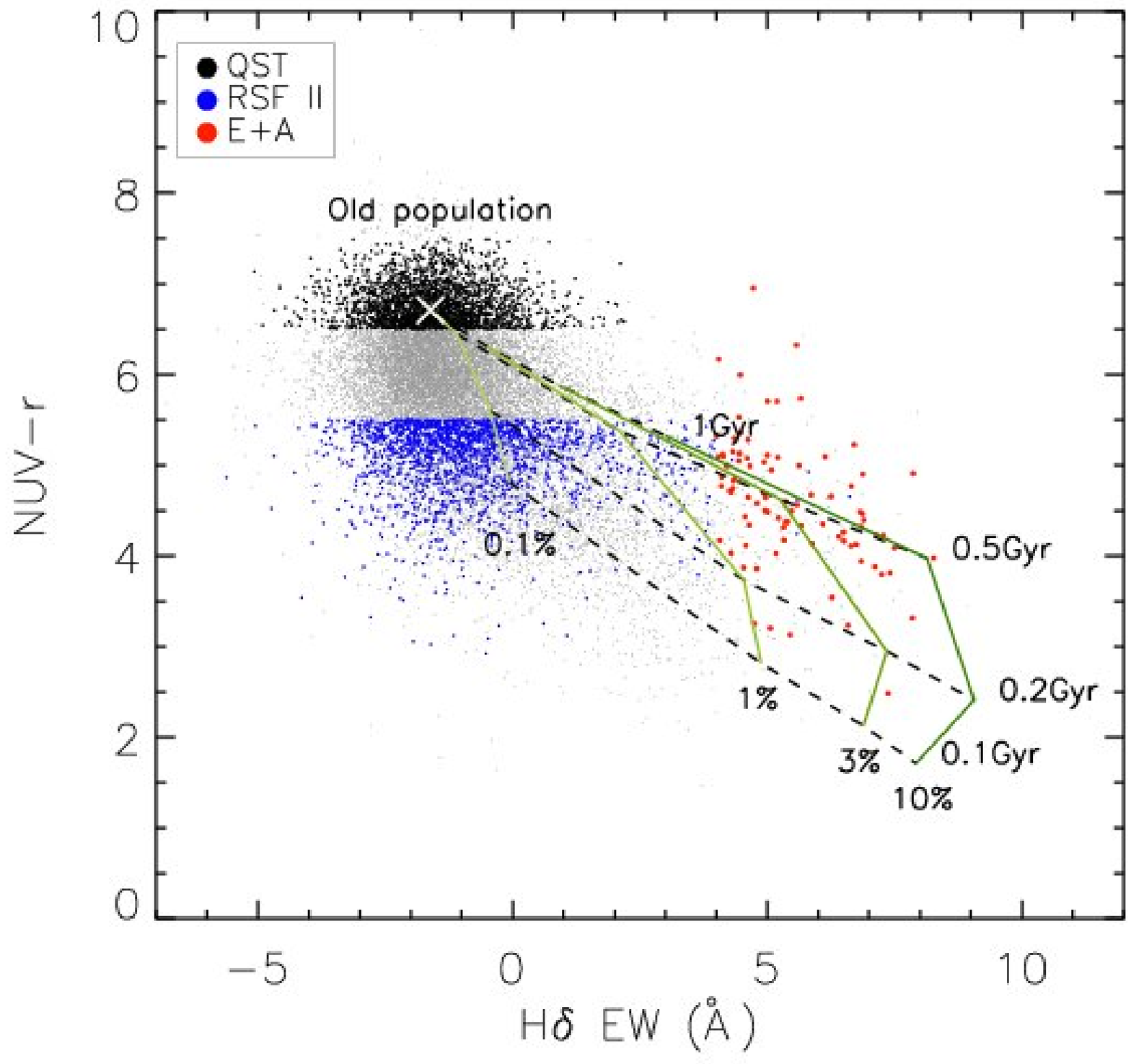}
\caption{H$\delta$ vs. $NUV-r$ for QST, RSF II, and E+A,
overlaid with the simple stellar population model grid. A median SED
of QST is used to represent the old population. We then add the young
population (Z = 0.02) with various combinations of age and mass
fraction to the underlying old population.}
\label{fig:bc03}
\end{center}
\end{figure*}
Figure~\ref{fig:bc03} shows possible evolutionary paths from E+A to QST via RSF
II by using a stellar population model \citep{bc03}.
We combine a representative old population (the median values of observed
H$\delta$ and $NUV-r$ of QST; large cross)
with a young population of solar metallicity ($Z$ = 0.02).
The stellar population model grids with added young population fractions of
$f$ = 0.1, 1, 3, and 10 per cent are plotted for the different ages
of $t$ = 0.1, 0.2, 0.5, and 1 Gyr at fixed metallicity of $Z$ = 0.02, as indicated by
the labels. The stellar population models provide two reasonable explanations
for the relationship among QST, RSF II, and E+A. If the star formation activities
were quenched quite a while ago, RSF II can be considered as the results of
passively evolved E+A (``the old E+As"). However, if star formation
was weak, RSF II represents a minor version of E+A (``the mild E+As").
We here refer to RSF II as ``E+a'', named after ``E+A''. The new acronym represents
elliptical galaxies (``E'') with a minority of A-type young stars (``a'').

Comparison of observational data with stellar population models suggests
two possible ways of connections between RSF II and E+A.
Although the origin of RSF II can be explained by both the ``old''
E+A scenario and the ``mild'' E+A scenario, E+A has an important property that
should not be overlooked. According to \citet{goto07}, E+As are very rare objects
in the lower-redshift Universe.
In the ``old'' E+A scenario, even though E+As are likely to evolve into RSF II with time,
there are not enough E+As to supply the entire RSF II population at any given point in time.
The scenario predicts that the mass distributions
of E+A, RSF II, and QST should be similar because the difference
among these species lies only on the recent star formation rate.
According to Figure~\ref{fig:mass}, however, E+A, RSF II, and QST have
a systematic difference in there masses and luminosities.
These indicate that a large fraction of RSF II galaxies should be explained
by the ``mild'' E+A scenario.

It is important to estimate the number fractions of RSF II galaxies
following two different paths. In Figure~\ref{fig:bc03}, the stellar population
model suggests that RSF II galaxies right below QST group (i.e. --3 \AA\, $\le$ H$\delta$ EW $<$ 2 \AA)
fit better into the ``mild'' E+A scenario. On the other hand, the rest of
RSF II galaxies located at the plume-like region toward E+As are
better explained by the ``old'' E+A scenario. Defining the old E+A regime as
2 \AA\, (the right edge of the mild E+A group) $\le$ H$\delta$ EW $\le$ 4 \AA\,
(the left edge of the E+A region), we estimate the number of RSF II galaxies
belonging to this group to be 103 ($\sim$ 5 per cent) among 1,936 RSF II galaxies.
We also estimate the number fraction of E+As in our sample. For a fair estimate
we only consider the number of E+As with GALEX UV information and the number
fractions of QST, RSF II, and E+As turn out to be 1.0 : 0.9 : 0.04. The number
of E+As ($\sim$ 5 per cent) is comparable to that of RSF II of the plume-like region,
further supporting their ``old'' E+A origin.
We therefore conclude that a large fraction of RSF II ($\sim$ 95 per cent) should be
the minor (i.e. weaker star formation) versions of E+A (i.e. mild E+A),
rather than passively evolved E+A (i.e. old E+A).

\section[]{Summary and Conclusion}
\label{sec:con}
The E+A galaxies may undergo the transition from
`blue cloud' to `red sequence' and eventually migrate to red sequence
early-type galaxies. An observational validation of this scenario
is to identify the intervening galaxy population that smoothly bridges the E+As to
red-sequence early-type galaxies.
In order to investigate the possible associations between E+As and
UV-excess galaxies, we have compared their Fundamental Planes.
We used the largest sample of 1,021 E+As selected from the SDSS DR6
and $\sim$ 20,000 early-type galaxies with GALEX UV data.
Based on recent star formation modes derived from $NUV-r$ colour and H$\alpha$ emission line,
we have broken early-type galaxies into QST (RSF-free), RSF I (ongoing weak SF),
RSF II (post weak star formation), and SH$\alpha$ (ongoing starburst) types.
The E+A FP is different from the QST FP, and this is
most likely due to a recent starburst in the central region
and their intrinsically smaller velocity dispersion.
Our sample indicates that RSF II galaxies
underwent starbursts that were weaker than those observed in E+As (``the mild E+A scenario'')
or the products of passively evolved E+As (``the old E+A scenario'').
The galaxies are characterized by UV-excess but no H$\alpha$ emission,
and this is essentially a conceptual generalisation of ``E+A'', in that
the Balmer absorption line in the ``E+A'' definition is replaced with
UV -- optical colours that are far more sensitive to RSF than the Balmer lines.
We refer to these UV-excess galaxies as ``E+a'' galaxies (named after ``E+A''),
which stands for elliptical (``E'') galaxies with a minority of A-type
(``a'') young stars. We suggest that most old ``E+A'' galaxies represent
the most recent arrivals to the red sequence.

A comparative study between E+a and E+A in terms of their preferred environment
will give additional insight into their relationship. For instance, E+As are preferentially
found in low-density environment in the nearby Universe and have an excess of neighbours
on small scale \citep{goto05}. A prediction from the present study is that the E+a population
would also share these properties with E+As. \citet{schawinski07} showed that the RSF fraction
is higher in the field environment, although they do not discriminate between RSF I and RSF II.
Further investigation is in progress on the environment of E+a and E+A galaxies in terms
of their local galaxy number density (Choi et al., in preparation).

\section*{Acknowledgments}
We thank the anonymous referee for careful reading of the manuscript
and the useful suggestions and comments.
Y. C. acknowledges the Department of Infrared Astrophysics of Institute of Space and
Astronautical Science (ISAS) for their hospitality. Especially, Y. C. is indebted to
Takao Nakagawa and Kumiko Nishimatsu for their support during her visit.
We thank H.Inami, H.Matsuhara, and K.Yun for useful discussion.
We would like to thank Changbom Park and Yun-young Choi for providing the catalogue of
morphologically-classified SDSS galaxies based on the DR4plus.
Y. C. acknowledges support from the Korea Research Foundation.

T. G. acknowledges support from the Japan Society for the Promotion of
Science (JSPS) through JSPS Research Fellowships for Young Scientists.
This work was supported by the Sasakawa Scientific Research Grant from
the Japan Science Society and by the Japan Society
for the Promotion of Science through Grant-in-Aid for Scientific Research 18840047.

S.-J. Y. acknowledges support from the Basic Research Program (grant No. R01-2006-000-10716-0)
and the Acceleration Research Program of the Korea Science and Engineering Foundation,
and from the Korea Research Foundation Grant funded by the Korean Government
(grant No. KRF-2006-331-C00134).

\label{lastpage}

\end{document}